# Geometric Correction and Mosaic Generation of Geo High Resolution Camera Images

Ankur Garg, Nitesh Thapa, Ghansham Sangar, Neha Gaur,
Meenakshi Sarkar, S. Manthira Moorthi, Debajyoti Dhar *

October 22, 2024

## Abstract

The Geo High Resolution Camera (GHRC) aboard the GSAT-29 satellite, operated by the Indian Space Research Organization (ISRO), is a cutting-edge 6-band Visible and Near Infrared (VNIR) imager positioned in geostationary orbit at 55°E longitude. It achieves a ground sampling distance of 55 meters at nadir and covers an instantaneous area of 110x110 km, capturing imagery of the entire Earth's disk thanks to its scan mirror mechanism. To comprehensively cover India, the GHRC employs a two-dimensional raster scanning technique, resulting in over 1,000 scenes that need to be skillfully stitched together into a seamless mosaic. This paper outlines the geolocation model and identifies potential sources of error affecting targeting accuracy. It also conducts an evaluation of the location accuracies. Challenges related to inter-band registration and inter-frame mosaicing with the GHRC are addressed. Several algorithms have been devised to address geometric correction, band-to-band registration, and seamless mosaic generation. Notably, adjustments to the instrument's interior alignment angles, guided by ground reference images during in-flight geometric calibration, have significantly improved pointing and location precision. A groundbreaking backtracking algorithm has been developed to rectify frame-to-frame mosaicing errors for large-scale mosaic generation. This algorithm harnesses a combination of geometric models, advanced image processing techniques, and space resection. As a result of these advancements, operational generation of full India mosaics with 100-meter resolution and high geometric fidelity is now a reality. These accomplishments mark substantial progress in the capabilities of the GHRC and hold profound implications for applications in Earth observation and monitoring.

GHRC, Geostationary orbit, Scan Mirror Geo-location model, Band to Band Misregistration, Targeting Accuracy, Location Accuracy, Frame Mosaicing, Space Resection.

## 1 Introduction

The Geo High Resolution Camera (GHRC) aboard the Indian GSAT-29 satellite is a remarkable technological achievement in the field of imaging payloads. It boasts an extraordinary imaging capability, achieving a ground sampling resolution finer than 55 meters at nadir. This places it among the highest spatial sampling cameras ever deployed in a geostationary orbit, standing alongside the renowned GeoFan4 [1]. The GHRC utilizes a mirror mechanism to scan its target, a technique that parallels the approach used by satellites such as the Indian National Satellite (INSAT) [2], the Geostationary Operational Environmental Satellite (GOES) [3] in geostationary orbits (GEO), and the Moderate Resolution Imaging Spectroradiometer (MODIS) [4] in lower earth orbits (LEO). These satellites employ linear detectors to gather data across various spectral channels, with the image's scanlines generated as the mirror rotates in the East-West direction. What distinguishes the GHRC is its utilization of a frame camera to capture snapshots, with spectral bands acquired through a filter wheel mechanism [5]. This innovative approach enables the snapshot camera to cover an instantaneous area of 110x110 km at nadir. For even larger coverage, the GHRC employs a raster scanning technique with a field-of-regard that encompasses the entire Earth disc. A critical component of the GHRC is its encoder, which plays a pivotal role in determining the position of the scan mirror. However, it's worth noting that this encoder possesses relatively lower noise fidelity. Consequently, it introduces system-level errors, which in turn present significant challenges in the geometric processing of the payload. To offer a comprehensive overview, Table 1 summarizes the key system parameters of the GHRC. This includes crucial specifications such as ground sampling resolution, coverage area, and other pertinent details that underscore the remarkable capabilities of this imaging payload. In summary, the GHRC represents a monumental leap in imaging technology, pushing the boundaries of what can be achieved from a geostationary orbit. Its innovative design and capabilities hold the potential to significantly advance Earth observation and monitoring applications.

---

*All authors are with Signal and Image Processing Group, Space Applications Center, Indian Space Research Organisation, Ahmedabad, India-380015

Table 1: GHRC System Parameters

| Parameter | Value | Remarks |
| --- | --- | --- |
| Spectral bands | 6 bands (460-870 nm) | Filter Wheel Based |
| Ground Sampling Distance(m) | 54 | For Indian region (from 55 to 85 m) |
| Swath (km) | 110x110 | In a single snap |
| Signal to Noise Ratio at 100 % albdeo | >100 | With Appropriate Exposure Setting |
| No. of scan axes | 2 (NS and EW) | |
| Minimum Step Size ($^0$) | 0.14(NS) and 0.07 (EW) | 20 % frame to frame overlap |
| Encoder Quantization (bits) | 21 | |
| Encoder Noise (P-P) (counts) | 5 | |

# 2 Geolocation Model

The geometric model for the GHRC is designed to calculate the precise ground location of each pixel based on a combination of telemetry data and calibrated instrument parameters. This model takes into account various frames of reference:

- **Focal Plane Frame:** This frame defines the look vector of each pixel at a specific moment in time. It essentially represents the orientation of the image in the instrument's focal plane.

- **Sensor Frame:** In this frame, the model establishes the orientation of the optical axis relative to a sensor reference cube. It provides a reference for the instrument's optical alignment.

- **Instrument Frame:** Within this frame, the model reflects the pixel's look vector with the scan mirror normal to obtain a vector in object space. This step is crucial for accurately determining the location on Earth's surface corresponding to each pixel.

  The instrument's geometric characteristics are assumed to be stable and have been calibrated. Parameters like the instrument's reference frame in relation to the platform, as well as the detectors' line of sight in reference to the instrument frame, are determined through in-orbit calibration. At a specific time:

- **Scan Mirror Frame:** This frame is determined by the orientation of the scan mirror angle relative to a cube representing the scan mirror. It defines how the scan mirror directs the incoming light.

- **Orbital Frame:** This frame is established by the satellite's ephemeris data. It provides the position and velocity information of the satellite at that moment.

- **Spacecraft Frame:** The spacecraft frame is defined by the attitude data. It specifies the orientation of the satellite in space.

The integration of these parameters culminates in the computation of the pixel's line of sight (LOS) within a fixed-earth frame. This LOS is critical in determining the exact point on Earth's surface where the pixel's imaging vector intersects. This intersection is calculated with respect to an earth ellipsoid, taking into account elevation data from a digital elevation model (DEM). This process yields a precise geographic location for each pixel. Regarding the spacecraft's orientation:

- The yaw axis, oriented towards the center of the Earth, serves as a pivotal reference point for the satellite's directional control.

- The pitch axis, orthogonal to both the yaw axis and the velocity vector, provides an orthogonal reference in relation to the satellite's trajectory and orientation in space.

- The roll axis, completing this triad, offers a perpendicular axis to the yaw and pitch, providing a complete three-dimensional orientation for the satellite.

Additionally, the model meticulously incorporates the scan mirror's position, which is precisely measured using an encoder. This positional data is then transformed into corresponding angles around the roll, pitch, and yaw axes. This information plays a critical role in precisely determining the orientation of the scan mirror, and consequently, the direction in which it steers incoming light.

To achieve an accurate and reliable assessment of GSAT-29's absolute three-axis attitude, data from two star sensors and three gyroscopes are harmoniously integrated. This fusion process is intricately engineered to deliver a robust and highly precise understanding of the satellite's orientation within the vast expanse of space.

In summary, this comprehensive geometric model stands as a pinnacle achievement in satellite imaging technology. It not only empowers the GHRC to capture high-resolution imagery but also enables the precise positioning and mapping of features on Earth's surface. This capability holds profound implications for a diverse range of applications, including environmental monitoring, disaster management, urban planning, and many more. The GHRC, equipped with its advanced imaging and precise geolocation capabilities, firmly establishes itself at the vanguard of Earth observation technology.

The look vector of a pixel is given by the column vector

$$u_{look} = [-1, tan(\psi_y), tan(\psi_z)]^T \quad (1)$$

where $\psi_y$ and $\psi_z$ are the look angles of the detector. The pre-launch measured alignment angles are used to rotate the look vector from focal plane frame to sensor frame and then finally to instrument frame using the following equation

$$u_{instr} = R^{instr}_{sensor} R^{sensor}_{focalplane} u_{look} \quad (2)$$

When the scan mirror is aligned parallel to yaw-pitch plane, the mirror normal is given by the column

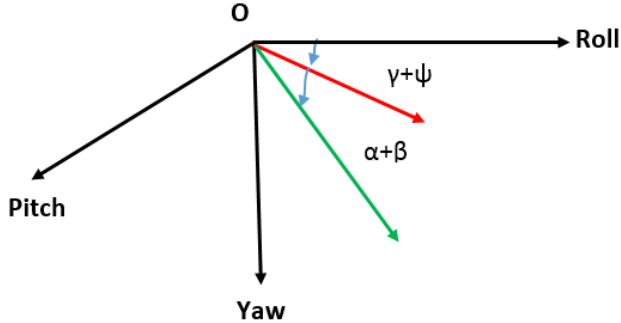

Figure 1: Scan Mirror Distortions Definitions

vector

$$n_{side} = [cos(\alpha + \beta)cos(\gamma + \psi), cos(\alpha + \beta) \\ sin(\gamma + \psi), sin(\alpha + \beta))]^T \quad (3)$$

where the angles $\alpha$, $\beta$, $\gamma$ and $\psi$ are as shown in Figure 1. $\alpha$ and $\gamma$ denote the mirror rotation axis being slightly tilted with respect to yaw-pitch plane. $\beta$ and $\psi$ define the wedge angles of the mirror surface which represent any mirror distortion. The effect of $\alpha$, $\beta$ and $\gamma$, $\psi$ on the ground is similar.

The scan mirror current orientation as given by the encoder is used to rotate the mirror normal to its current pointing location. The scan mirror normal is then rotated to instrument frame using the pre-launch alignment matrix having mounting angles information of scan mirror cube w.r.t instrument. The current mirror normal in instrument frame is given by

$$n_{instr} = R_{mirror}^{instr} R_{mirr} n_{side} \tag{4}$$

The unit vector of the mirror surface normal is given by

$$\hat{n_{instr}} = n_{instr}/|n_{instr}| \tag{5}$$

The image viewing vector $u_{instr}$ is reflected off the mirror to get the viewing vector in the object space using the following equation.

$$u_{instr}^{obj} = u_{img} - 2\hat{n_{instr}}(u_{img}.\hat{n_{instr}}) \tag{6}$$

At any given time, the satellite roll ($\xi_r$), pitch ($\xi_p$) and yaw ($\xi_y$) angles are used to go from spacecraft to orbital frame of reference. The orbital frame is defined by the position and velocity vectors and the directional cosine matrix is used to convert the orbital frame to Earth Centered Inertial (ECI) frame of reference. The transformation relation between ECI and ECER (Earth Centered Earth Rotating) is given by the sidereal angle. The overall transformation is given by

$$R_{instr}^{ecr} = R(t)_{eci}^{ecr} R_{orb}^{eci} \tag{7}$$

The look vector in the ECER frame is

$$u_{ecr} = R_{ecr}^{instr} u_{instr}^{obj} \tag{8}$$

The intersection of the line-of-sight with the WGS-84 ellipsoid is then calculated as:

$$x_{ellipse} = p_{ecr} + \lambda u_{ecr} \tag{9}$$

where $\lambda$ is the scale factor and $p_{ecr}$ is the position of the satellite. The ellipsoid intersection is calculated by turning the problem into a unit intersection problem. After computing the latitude, longitude at a given height for every pixel, data is uniformly map projected into Lambert conformal conic projection (LCC) with cubic interpolation.

## 3 Error Sources and Related Issues

The geometric error sources in GHRC are mainly either due to platform or occur from within the instrument. The factors which affect platform performance are (a) Orbit Determination inaccuracy, (b) Attitude Determination inaccuracy, (c) Thermo-elastic deformations which can lead to change in alignment of star sensors, gyroscopes with respect to payload (d) Time synchronization errors, (e) Gyro drift, scale and misalignment errors, (f) Control accuracy, (g) Noise in the knowledge of scan mirror position, and (h) any static biases occurring due to lack of knowledge of the mounting angles. The GHRC scan mirror operates in a closed loop manner in which it fluctuates around the commanded value within a pre-defined threshold. This value has been set as 15 encoder counts which corresponds to 60 pixels on the ground in EW and 30 pixels in NS directions. Setting the value to a lesser threshold makes the scan mirror take more time to settle down and vice-versa. The peak to peak mirror encoder noise is expected to be ±5 counts which amounts to 40 pixels difference in EW and 20 pixels difference (end to end) between adjacent frames. Following are the expected issues in geometric processing of GHRC due to the above sources of errors.

**Targeting Error :** The targeting error for any system is the difference between its intented location and the location captured. All the above listed sources of error affect the targeting performance of GHRC. The worst case targeting accuracy was expected to be $\sim 0.15^0$ in roll & pitch direction and $\sim 0.20^0$ in yaw direction which is equivalent to $\sim 93$km and $\sim 125$km respectively on the ground.

**Location Error :** The location error of an imaged ground point is the difference between its real location and the location computed from telemetry data (orbit, attitude) and calibration outcomes (line of sight), without geometric model refinement with GCPs. Location accuracy is dependent on the precise knowledge of various sources and hence will suffer if the knowledge is improper. In GHRC, the worst case location accuracy was expected to be $\sim 0.05^0$ in all the three axis which is equivalent to

∼31km on the ground.

**Band's Misregistration :** Given a ground point P, the difference in true projection of P in 2 arbitrary bands of the same instrument is known as band to band misregistration (BBR). Like the location accuracy, BBR arises due to change in system parameters when different bands of the same scene are getting acquired. Misregistration is caused by relative drift in the platform during the frame acquisitions and fluctuation of the mirror around the intended values.

**Mosaic Errors :** Because of the two dimensional scanning mechanism of GHRC to acquire larger areas, each frame on the ground has 8 neighbouring frames with overlaps. Error in the overlapping region in any of these frames leads to geometric non conformality in the mosaic. Scan mirror encoder noise is the major contributor for mosaicing errors across frames. Lack of knowledge of platform drift between the frame acquisitions will also lead to this error.

## 4 Error Analysis and Correction Methodology

During the commissioning phase of GHRC, analysis of various sources of errors is done to validate the overall performance of the system and algorithms were built to correct for the following errors.

**Band-Band Misregistration(BBR) :** As mentioned in the section III, band to band misregistration is mainly caused by fluctuation of the scan mirror around the intended value and platform drift. To validate the platform drift of GSAT-29, GHRC was operated in video mode where the scan mirror was forced to point at a single place and multiple frames were acquired. Figure 2 shows the deviation in angles about roll and pitch directions observed during this exercise. The observed high frequency fluctuation can be attributed to that coming from the scan mirror and the mean deviation is coming from the platform. Ignoring the high frequency deviations, the mean drift value was found to be of the order of $10^{-5} deg/sec$ in both roll and pitch directions which corresponds to around 125m (∼2.27 pixels) on the ground in 20sec, during which all the bands of the scene are acquired. This shows that the major contributor of BBR are fluctuations coming from the scan mirror. Figures 3 & 4 show the histogram of the BBR induced w.r.t red channel for a 30x30 scan (4500 frames). As expected the values are within 30 pixels for roll and within 60 pixels in pitch directions coming from the scan mirror fluctuations. To correct for misregistration, either system level knowledge with the model described in section II or an image based correction which does not require any knowledge of the system parameters can be used. For system level correction, precise knowledge of the scan mirror encoder angles is needed. Due to error/noise in knowledge of the encoder values, this approach lead to large residues in BBR and hence is not being used. The error induced due to scan mirror can be approximated by shift in the location of ground feature along line and pixel directions. Image based normalized cross correlation is used to estimate and correct this affect. For each band, the scan and pixel shifts are calculated at subpixel level by fitting a two-dimensional polynomial over the correlation surface with respect to red channel and the data is resampled to compensate the shifts. Figure 5 show sample images before and after BBR correction with image based approach. The BBR was evaluated for multiple full India scans and the performance is found to be better than a 0.25 pixel with 98% circular error(CE) using this approach as shown in Figures 3 & 4.

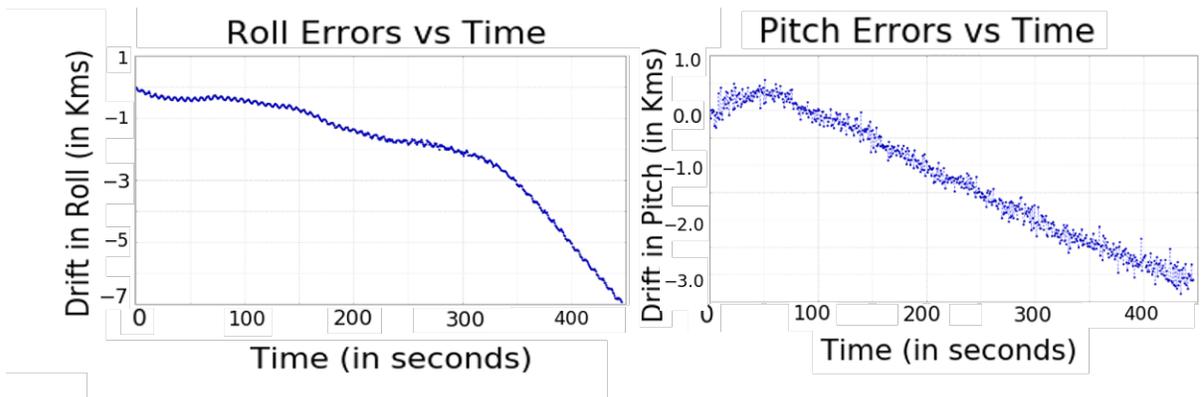

Figure 2: Stability of GSAT-29

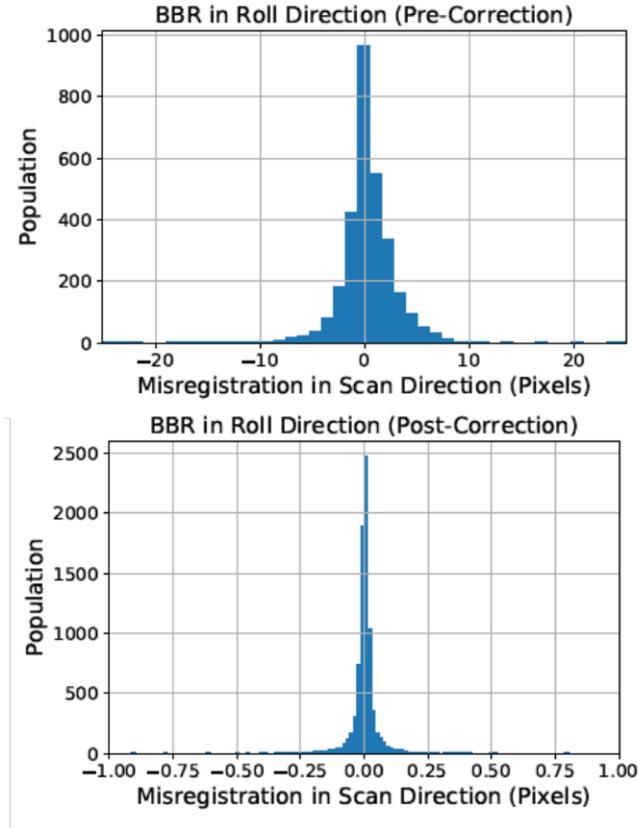

Figure 3: BBR in Roll Direction

**Targeting Accuracy :** In the initial phase operations, the targeting error was found to be around ∼200km in Northing and ∼100km in Easting direction. The major error source affecting targeting error in GHRC was found to be mounting errors related to the instrument. Based on automatic ground control points estimated from Landsat Enhanced Thematic Mapper(ETM) reference images using normalized cross correlation, the alignment angles of the scan mirror normal wrt scan mirror cube normal were modified. Table 2 shows the alignment angles before and after the bias updation. With the new alignment angles, the targeting accuracy was monitored for further dates and it was found that the targeting performance improved to ∼30km during the nominal imaging operations.

Table 2: GHRC Alignment Angles

| Parameter | Pre-Launch | After Calibration |
|---|---|---|
| $\text{EW}_{RefAngles}(deg)$ | 195.44 | 195.60 |
| $\text{NS}_{RefAngles}(deg)$ | 15.79 | 15.20 |

**Location Accuracy :** The mounting error related to instruments also lead to location inaccuracies ∼200km in Northing and ∼100km in Easting direction in the initial phase. After the alignment angles were modified, the location inaccuracy came down to ∼15km for further dates.

**Mosaic Generation :** As mentioned in the section III, frame to frame mosaicing error is mainly affected by scan encoder error when system level correction is performed. Frame mosaicing is a well known research problem in computer vision domain. Many algorithms[6], [7] for Unmanned Aerial Vehicle(UAV) and airborne scene mosaicing have been developed by researchers. All these depend heavily on the correspondence points in the overlapping regions of the scenes and on bundle block adjustment. In GHRC, the problem of full India scene mosaicing is even more challenging as the number of scenes are large (∼ 1 thousand) and also due to presence of large amount of ocean with small amount of land region in the lower part of India which make finding correspondence difficult. To correct for the same and generate large scale mosaics, a backtracking algorithm incorporating geometric model, image based techniques with well known photogrammetry concept of space resection has been devised. Space resection is the problem of finding or modifying the interior/exterior orientation of the scene with GCPs. Let us assume that the ground projection of the frame is given by model $f$. The model is parameterized by many variables as defined in section II. To find or update any of the model parameters, sum of square errors between the actual and the desired point is minimized.

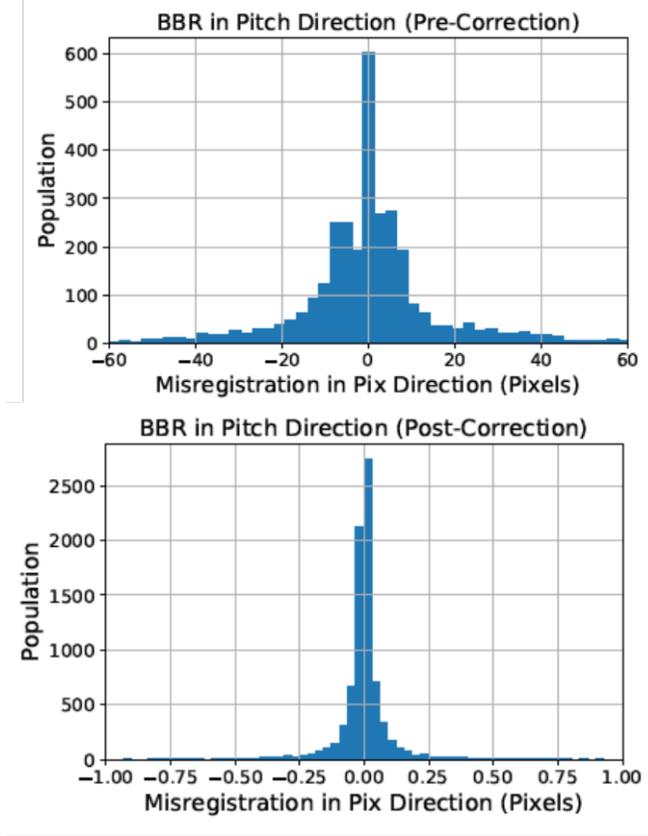

Figure 4: BBR in Pitch Direction

$$S(\beta) = \sum_{i=1}^{m}(y_i - f(x_i, \beta)) \tag{10}$$

By linearizing the non-linear function $f$ by Taylor series expansion,

$$S(\beta + \delta\beta) = \sum_{i=1}^{m}(y_i - (f(x_i, \beta) + \delta\beta.\frac{\partial f(x_i, \beta)}{\partial \beta}))^2 \tag{11}$$

Setting $\frac{\partial S(\beta+\delta\beta)}{\partial \delta\beta} = 0$, we get

$$(J^T J)\delta\beta = J^T(y - f(\beta)) \tag{12}$$

$$\delta\beta = (J^T J)^{-1}(J^T(y - f(\beta))) \tag{13}$$

where $J = [\frac{\delta f}{\delta b1}, \frac{\delta f}{\delta b2}, ...]$ and $b1, b2$ are the parameters of the model to be updated. The above equation is solved using gradient descent method. Using space resection, either all or selective model parameters can be updated. We modify only the scan mirror cube to instrument cube roll and pitch angles after estimating correspondences in overlapping areas of the adjacent frames. The overlap between the frame is not static and changes dynamically with encoder counts and satellite attitude. The approximate overlapping portion is decided dynamically based on current scan mirror encoder angles and attitude of the satellite according to following equation.

$$\begin{aligned}over_{pr} = (1.0 - 2.0 * abs(mPitch_{currF} - mPitch_{prF})/0.176 \\ + tan(abs(pPitch_{currF} - pPitch_{prF})) * 35786.0/110.0\end{aligned} \tag{14}$$

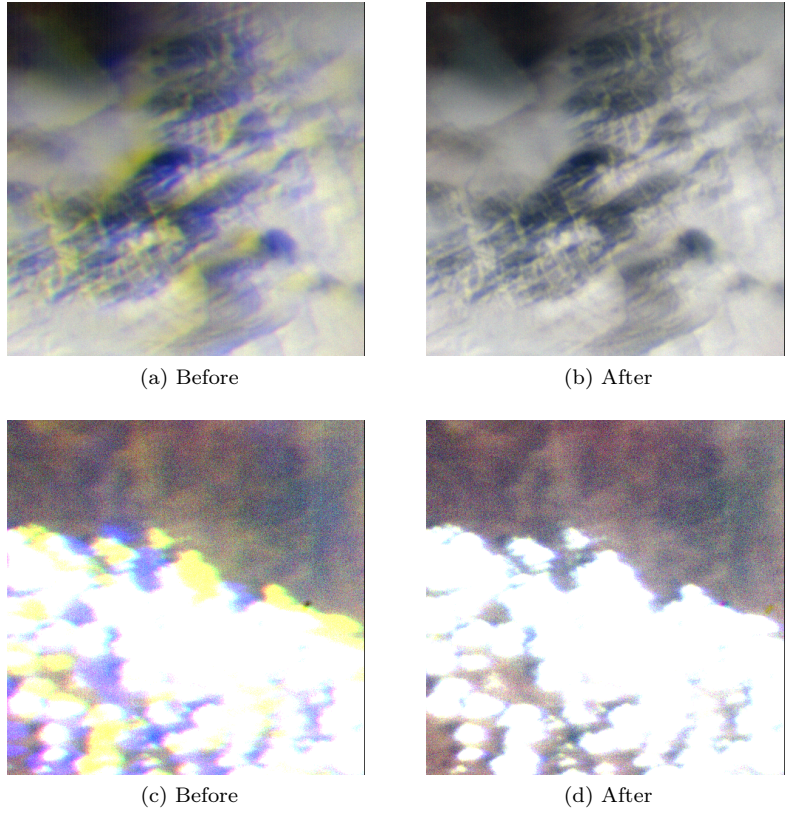

Figure 5: Before and After BBR Correction

$$over_{up} = (1.0 - 2.0 * abs(mRoll_{currF} - mRoll_{upF})/0.176 \\ + tan(abs(pRoll_{currF} - pRoll_{upF})) * 35786.0/110.0 \quad (15)$$

where $over_{pr}$ and $over_{up}$ are the overlapping values with respect to previous frame and frame lying above the current frame. $mPitch_{currF}$, $mPitch_{prF}$, $mRoll_{currF}$, $mRoll_{prF}$ are the pitch and roll angles coming from the scan mirror rotation for current and previous frames respectively. $pRoll_{currF}$, $pRoll_{prF}$, $pPitch_{upF}$ and $pPitch_{upF}$ are the platform roll and pitch angles for current and the frame lying above the current frame respectively. All the angles are in degrees. $0.176^0$ is the Field of View(FOV) of the frame, 35786.0 is the altitude of the satellite and 110 km x 110 km is the Instantaneous Ground Field of View(IGFOV) of the frame. The basic idea is that if all the frames are aligned with respect one of their neighbouring frames in overlapping regions, then there would not be any mosaicing errors. The algorithm for correcting the mosaicing error is a two-pass algorithm and relies on backtracking. In the first pass, for each frame, one of the neighbouring frame to which the current frame is to be aligned is selected. In the second pass, the value of the model parameters is modified and then final georeferencing is done to build the mosaic. In this way, the final corrected mosaic is created using only one resampling operation. The detailed approach is described in Algorithm 1 and 2. Based on whether the scan is happening towards east (right) or west (left) in the two-dimensional GHRC scanning, first the previous frame is selected with respect to current frame and phase correlation [8] is done on the overlapping region. If the confidence in the estimated value is acceptable, then we take the previous frame as reference for the current frame and move ahead to the next frame. The correlation value is used as the measure of confidence. The first peak value is compared to the second peak and if the first peak is atleast 50% larger than the second peak then there is enough confidence in the estimate. The issue arises when the correct shifts cannot be estimated with good confidence due to either the feature in the overlapping region being too homogeneous or cloudy. In such cases, either the frame lying above the current frame should be used to calculate the mosaic error or we move ahead to the next frames for the current moment, and backtrack whenever a frame in future is found which has good confidence with the frame above it. In this way, all consecutive frames which cannot be correlated with respect to their previous frame or to the frame lying above are stored and revisited during backtracking whenever a future frame is found which has good correlation with respect to frame lying above it. Appropriate flag for previous, top and opposite is set. If all three conditions are not met, then nothing can be done for the current frame and it will be generated with only system knowledge. After Algorithm 1 is completed, result is a reference frame corresponding to each frame. The reference frame for a frame is used for correlating and modifying the model parameters in Algorithm 2 for that frame. In Algorithm 2 for each frame, the algorithm checks if the flags related to neighbour, up and opposite are false, then the frame has to be generated with system knowledge. If the neighbour flag is true, then the overlap with the neighbour is established as described by equations 14 and 15, and the model parameters are modified with space resection. Similar procedure is performed if up flag is true. If opposite flag is true, then all the consecutive frames with opposite true flag are counted and the frame parameters are adjusted moving in opposite direction. After algorithm 1 and 2 are completed, we should get a seamless mosaiced image with very only few localized errors at places where all the conditions were not met. Figures 6 show the mosaic errors occurring when the frames are georeferenced and mosaiced only with the system knowledge. Figure 7 show the images before and after mosaic errors correction using our approach. A seamless full India mosaic generated after processing with the above algorithm is shown in Figure 8.

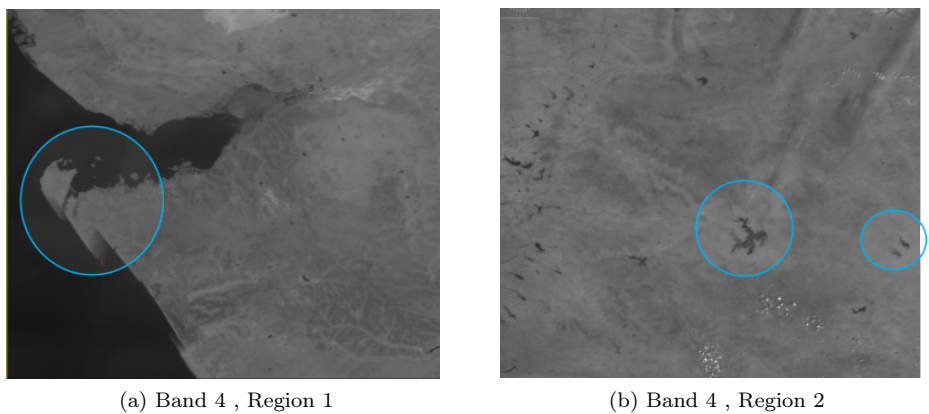

(a) Band 4 , Region 1    (b) Band 4 , Region 2

Figure 6: Mosaic Errors

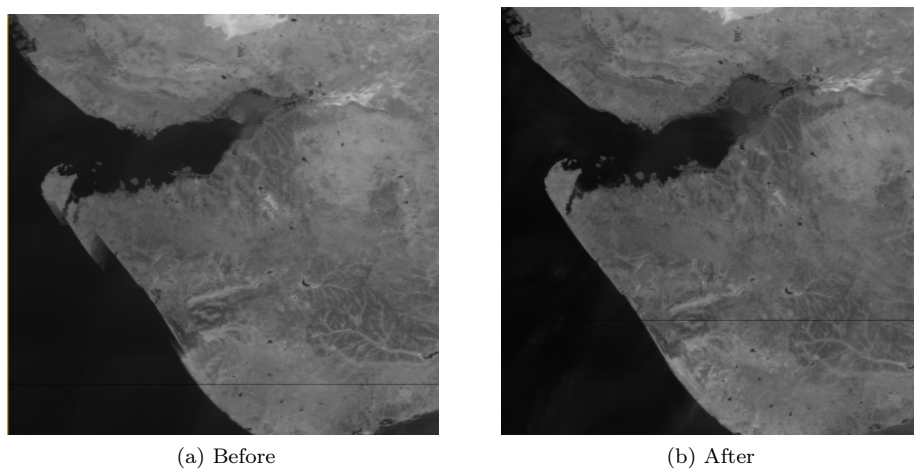

(a) Before    (b) After

Figure 7: Mosaic Error Correction

**Result:** Reference Frame Selection
**Initialization** :
takeNeighbour[#frames] = true
takeUp[#frames] = false
takeOpposite[#frames] = false ;
**while** *All the frames are not finished* **do**
    Let the current frame be i;
    **if** *Scanning Left/Right* **then**
        find overlap with frame right/left to i;
        estimate the shifts with phase correlation in the overlapping region;
        **if** *Shift not estimated with confidence* **then**
            takeNeighbour[i] = false;
            find overlap with frame upwards to i;
            estimate the shifts with phase correlation;
            **if** *Shift estimated with confidence* **then**
                takeUp[i] = true;
            **end**
        **end**
        **if** *(!takeNeighbour[i] && !takeUp[i])* **then**
            oppositeFrame=1;
        **end**
        **if** *(oppositeFrame>0)* **then**
            **if** *(takeUp[i] == true)* **then**
                counter=0;
                **while** *(oppositeFrame>1)* **do**
                      thisFrame=i-counter;
                      find overlap with frame right/left to i;
                      **if** *Shift not estimated with confidence* **then**
                          break loop;
                      **else**
                          takeUp[thisFrame]=false; takeNeighbour[thisFrame]=false;
                          takeOpposite[thisFrame]=true;
                      **end**
                    oppositeFrame- -;
                    count++;
                **end**
                oppositeFrame=0;
            **else**
                oppositeFrame++;
            **end**
        **end**
    **end**
**end**
**Algorithm 1:** Backtracking Algorithm for finding Reference Frames to be used in Algorithm 2

**Result:** Create Mosaic Image
**while** *All the frames are not finished* **do**
 Let the current frame be i;
 **if** *Scanning Left/Right* **then**
  **if** *(!takeNeighbour[i] && !takeOpposite[i] && !takeUp[i])* **then**
   skip this frame;
  **end**
  **if** *(takeNeighbour[i] == true)* **then**
   find overlap with frame right/left to i;
   Perform resection and update parameters for i;
  **end**
  **if** *(takeUp[i] ==true)* **then**
   find overlap with frame upward to i;
   Perform resection and update parameters for i;
  **end**
  c=0;
  //count the number of frames to be corrected in opposite direction;
  **if** *(takeOpposite[i]==true)* **then**
   **while** *takeOpposite[i+c]* **do**
    c++;
   **end**
  **end**
  for each of the opposite frame (i-c) till the current frame i, find the overlap region towards right/left with respect to their neighbors and perform resection;
 **end**
 Georeference current frame with the model described in Section II;
**end**

**Algorithm 2:** Mosaic Generation Algorithm

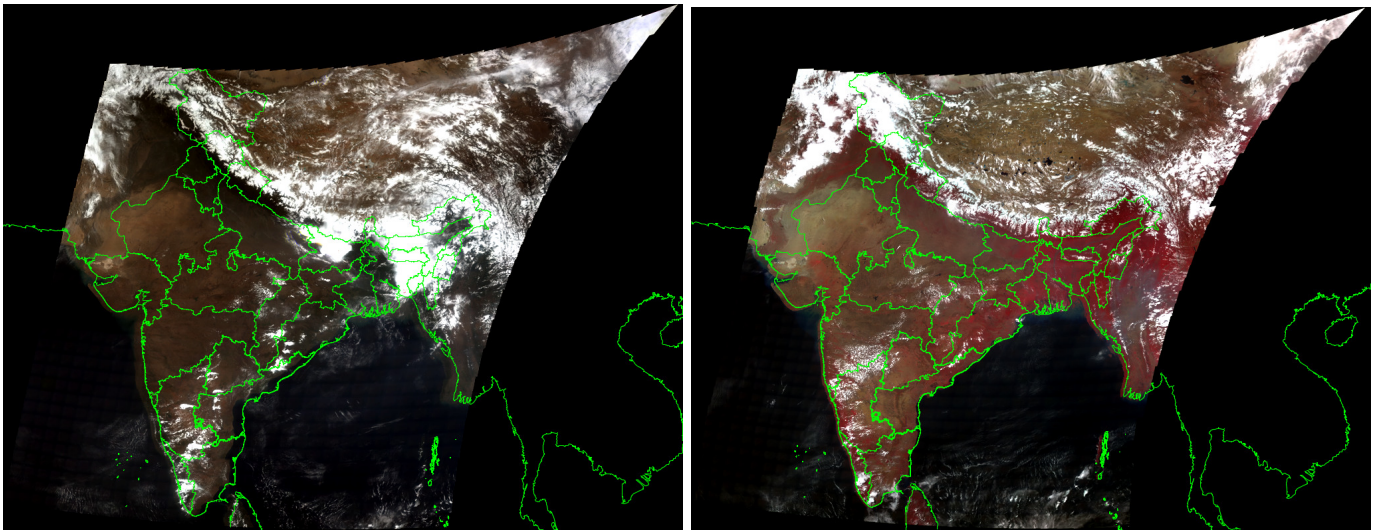

(a) False Colour Composite     (b) Natural Colour Composite

Figure 8: Full India Mosaic Overlayed with Vector Layer

# 5　Conclusion

This paper describes the error sources affecting targeting, location accuracies, inter-band registration and inter-frame mosaicing of GHRC. Algorithms for geometric correction, band to band registration and seamless mosaic generation have been desribed. For correcting the mosaic errors, a hybrid approach incorporating system knowledge and image processing approach has been developed and decribed. The approaches described here can be used to solve similar problems occurring in other geostationary satellites.

# Acknowledgment